\documentclass[aps,prl,reprint,superscriptaddress]{revtex4-1}
\usepackage{graphicx}
\usepackage{epstopdf}
\usepackage{dcolumn}
\usepackage{gensymb}
\usepackage{booktabs}
\usepackage{dcolumn}
\newcolumntype{x}{D{.}{.}{6.6}}
\newcolumntype{y}{D{.}{.}{5.6}}
\newcolumntype{z}{D{.}{.}{5.7}}
\newcolumntype{f}{D{.}{.}{7.9}}
\newcolumntype{e}{D{.}{.}{6.8}}
\usepackage[centerlast]{caption}
\usepackage{bm}        
\usepackage{amssymb,amsmath}   
\usepackage{subfig}

\begin{document}


\title{Isomer shift and magnetic moment of the long-lived 1/2$^{+}$ isomer in $^\text{79}_\text{30}$Zn$_{49}$: \\
        signature of shape coexistence near $^\text{78}$Ni}


\author{\mbox{X.F. Yang}}
\email{xiaofei.yang@fys.kuleuven.be}
\affiliation{KU Leuven, Instituut voor Kern- en Stralingsfysica, B-3001 Leuven, Belgium}
\author{\mbox{C. Wraith}}
\affiliation{Oliver Lodge Laboratory, Oxford Street, University of Liverpool, L69 7ZE, United Kingdom}
\author{\mbox{L. Xie}}
\affiliation{School of Physics and Astronomy, The University of Manchester, Manchester M13 9PL, United Kingdom}
\author{\mbox{C. Babcock}}
\affiliation{Oliver Lodge Laboratory, Oxford Street, University of Liverpool, L69 7ZE, United Kingdom}
\affiliation{EN Department, CERN, CH-1211 Geneva 23, Switzerland}
\author{\mbox{J. Billowes}}
\affiliation{School of Physics and Astronomy, The University of Manchester, Manchester M13 9PL, United Kingdom}
\author{\mbox{M.L. Bissell}}
\affiliation{School of Physics and Astronomy, The University of Manchester, Manchester M13 9PL, United Kingdom}
\affiliation{KU Leuven, Instituut voor Kern- en Stralingsfysica, B-3001 Leuven, Belgium}
\author{\mbox{K. Blaum}}
\affiliation{Max-Plank-Institut f\"{u}r Kernphysik, D-69117 Heidelberg, Germany}
\author{\mbox{B. Cheal}}
\affiliation{Oliver Lodge Laboratory, Oxford Street, University of Liverpool, L69 7ZE, United Kingdom}
\author{\mbox{K.T. Flanagan}}
\affiliation{School of Physics and Astronomy, The University of Manchester, Manchester M13 9PL, United Kingdom}
\author{\mbox{R.F. Garcia Ruiz}}
\affiliation{KU Leuven, Instituut voor Kern- en Stralingsfysica, B-3001 Leuven, Belgium}
\author{\mbox{W. Gins}}
\affiliation{KU Leuven, Instituut voor Kern- en Stralingsfysica, B-3001 Leuven, Belgium}
\author{\mbox{C. Gorges}}
\affiliation{Institut f\"{u}r Kernphysik, TU Darmstadt, D-64289 Darmstadt, Germany}
\author{\mbox{L.K. Grob}}
\affiliation{Experimental Physics Department, CERN, CH-1211 Geneva 23, Switzerland}
\affiliation{Institut f\"{u}r Kernphysik, TU Darmstadt, D-64289 Darmstadt, Germany}
\author{\mbox{H. Heylen}}
\affiliation{KU Leuven, Instituut voor Kern- en Stralingsfysica, B-3001 Leuven, Belgium}
\author{\mbox{S. Kaufmann}}
\affiliation{Institut f\"{u}r Kernphysik, TU Darmstadt, D-64289 Darmstadt, Germany}
\affiliation{Institut f\"{u}r Kernchemie, Universit\"{a}t Mainz, D-55128 Mainz, Germany}
\author{\mbox{M. Kowalska}}
\affiliation{Experimental Physics Department, CERN, CH-1211 Geneva 23, Switzerland}
\author{\mbox{J. Kraemer}}
\affiliation{Institut f\"{u}r Kernphysik, TU Darmstadt, D-64289 Darmstadt, Germany}
\author{\mbox{S. Malbrunot-Ettenauer}}
\affiliation{Experimental Physics Department, CERN, CH-1211 Geneva 23, Switzerland}
\author{\mbox{R. Neugart}}
\affiliation{Max-Plank-Institut f\"{u}r Kernphysik, D-69117 Heidelberg, Germany}
\affiliation{Institut f\"{u}r Kernchemie, Universit\"{a}t Mainz, D-55128 Mainz, Germany}
\author{\mbox{G. Neyens}}
\affiliation{KU Leuven, Instituut voor Kern- en Stralingsfysica, B-3001 Leuven, Belgium}
\author{\mbox{W. N\"{o}rtersh\"{a}user}}
\affiliation{Institut f\"{u}r Kernphysik, TU Darmstadt, D-64289 Darmstadt, Germany}
\author{\mbox{J. Papuga}}
\affiliation{KU Leuven, Instituut voor Kern- en Stralingsfysica, B-3001 Leuven, Belgium}
\author{\mbox{R. S\'{a}nchez}}
\affiliation{GSI Helmholtzzentrum f\"{u}r Schwerionenforschung, D-64291 Darmstadt, Germany}
\author{\mbox{D.T. Yordanov}}
\affiliation{Institut de Physique Nucl\'eaire, CNRS-IN2P3, Universit\'e Paris-Sud,Universit\'e Paris-Saclay, 91406 Orsay, France}


\date{\today}
\begin{abstract}
Collinear laser spectroscopy has been performed on the $^{79}_{30}$Zn$_{49}$ isotope at ISOLDE-CERN. The existence of a long-lived isomer with a few hundred milliseconds half-life was confirmed, and the nuclear spins and moments of the ground and isomeric states in $^{79}$Zn as well as the isomer shift were measured. From the observed hyperfine structures, spins $I = 9/2$ and $I = 1/2$ are firmly assigned to the ground and isomeric states. The magnetic moment $\mu$ ($^{79}$Zn) = $-$1.1866(10) $\mu_{\rm{N}}$, confirms the spin-parity $9/2^{+}$ with a $\nu g_{9/2}^{-1}$ shell-model configuration, in excellent agreement with the prediction from large scale shell-model theories. The magnetic moment \mbox{$\mu$ ($^{79m}$Zn) = $-$1.0180(12) $\mu_{\rm{N}}$} supports a positive parity for the isomer, with a wave function dominated by a 2h-1p neutron excitation across the $N = 50$ shell gap. The large isomer shift reveals an increase of the intruder isomer mean square charge radius with respect to that of the ground state: $\delta \langle r^{2}_{c}\rangle^{79,79m}$ = +0.204(6) fm$^{2}$, providing first evidence of shape coexistence.
\end{abstract}



\maketitle

The nuclear shell model, based on experimental observations near the valley of stability, has guided our studies and understanding of the nuclear quantum many-body system for decades. With the development of accelerators and isotope separators, exotic isotopes far from stability became accessible experimentally, revealing interesting phenomena, such as shape coexistence near closed shells and reduction in the energy of shell gaps as a function of proton or neutron number. Those observations challenge the persistence of the established \lq\lq magic numbers", and at the same time, drive the improvement of various theoretical models. The interplay between the stabilizing effect of a closed shell on one hand and the residual interactions between protons and neutrons outside closed shells on the other hand, leads to the concept of \lq shape coexistence\rq, where normal near-spherical and deformed structures coexist at low energy \cite{Heyde1983,Heyde2011}. The deformed structures appear as multi-particle multi-hole (mp-mh) excitations of protons or neutrons across a \lq\lq closed" shell gap \cite{Heyde2011}. These so-called \lq intruder'  configurations become low in energy in isotopes with a nearly closed neutron (or proton) shell, and with the partner nucleon having a near-mid-shell nucleon number. Their excitation energy is strongly reduced due to the gain in correlation energy in these mid-shell regions. These deformed intruder states even become the ground state in regions where the shell gap is reduced (e.g. in the island of inversion \cite{Neyens2011}). In the recent review by Heyde and Wood \cite{Heyde2011}, an overview is given in Fig. 8 of regions where shape coexistence has been established. Shape coexistence at low energy appears all over the nuclear chart, always along closed shells and away from the doubly-magic isotopes. For example, along proton closed shells $Z$ = 82 and $Z$ = 50, the shape coexistence has been well established by a combination of various experimental techniques \mbox{\cite{Lead2000,Kluge2003,Bree2014}}. In the case of neutron-closed shells, experimental evidence for shape coexistence has been observed along $N = 20$ \cite{32MgWimmer2010,34SiRotaru2012, Neyens2015, Gade2016}, along $N = 28$ \cite{44SForce2010,43SGaudefroy2009,43SChevrier2012} as well as along the sub-shell gap $N$ = 40 \cite{Gade2016}.

Along the next neutron closed shell $N$ = 50, experimental evidence of shape coexistence has not been reported so far, although intruder mp-mh states have been observed in some $N$ = 49 isotones near the proton mid-shell around $Z$ = 34 \cite{Heyde1983}. Low-lying intruder states with spins 1/2$^{+}$ and 5/2$^{+}$ were first seen in $^{83}_{34}$Se via the (d,p) transfer reaction \cite{Lin1965,Mon1978}, and subsequently discussed in the decay spectroscopy work of Meyer $et$ $al$. \cite{Meyer1982}. Similar intruder states were also observed in the $^{81}_{32}$Ge isotope \cite{Hoff1981}, one of which is an unexpected long-lived $1/2^{+}$ isomeric state, although a 1/2$^{-}$ isomer with a $\nu 2p_{1/2}^{-1}$ configuration exists in the heavier $N$ = 49 isotones ($^{83}_{34}$Se, $^{85}_{36}$Kr, $^{87}_{38}$Sr). In the latter isotones, the intruder 1/2$^{+}$ level appears at a higher energy and its excitation energy decreases from $Z$ = 38  to $Z$ = 34 \cite{Meyer1982}. In $^{81}$Ge, this level sits below the normal 1/2$^{-}$ level and therefore is isomeric \mbox{(half-life: 7.6 s)}, indicating a possible shape coexistence in this region \cite{Heyde1983}. Early theoretical calculations in a Coriolis-coupling model by Heller $et$ $al$. \cite{Heller1974} showed that a deformation of $\beta \sim 0.2$ is needed to explain low energy levels in $N$ = 49 nuclei. Recently, a transfer reaction study reported on the existence of positive-parity levels in $^{79}$Zn \cite{Orlandi2015}, one of which is isomeric (half-life was not measured) with a tentative spin/parity of 1/2$^{+}$.
\begin{table*}[t!]
\vspace{-3mm}
\caption{\footnotesize{Hfs constants for $^{79,79m}$Zn and  $^{67}$Zn and the extracted nuclear moments relative to the high-precision values of $^{67}$Zn \cite{Ad67Zn,
67Zn-moment,Q67Zn}.}}\label{moment}
\vspace{-3mm}
\renewcommand*{\arraystretch}{1.0}
\begin{tabular}{ l  l  f f  f  f   e  e}
\hline\hline
A &
                    \multicolumn{1}{c}{$I^{\pi}$}&
                   \multicolumn{1}{c} {$A(^{3}S_{1})$ }&
                   \multicolumn{1}{c} {$A(^{3}P_{2})$ }&
                   \multicolumn{1}{c} {$B(^{3}P_{2})$ }&
                   \multicolumn{1}{c} {$\mu_{\rm{exp}}$}&

                   \multicolumn{1}{c} {$Q_{\rm{s,exp}}$}
                   \\
      &
      &
      \multicolumn{1}{c}{(MHz)}                      &
      \multicolumn{1}{c}{(MHz)}                       &
      \multicolumn{1}{c}{(MHz)}                      &
      \multicolumn{1}{c}{($\mu_{\rm{N}}$)}           &
         \multicolumn{1}{c}{($b$)}  \\
\hline
67 & $5/2^{-}$ &+1267.5(10)\footnotemark [1] &+531.987(5)\footnotemark[2] &+35.806(5)\footnotemark[2] &+0.875479(9)\footnotemark[3] &  +0.150(15)\footnotemark[4]\\
67 & $5/2^{-}$ &+1266.5(18) &+531.2(11)  &+40.9(72)  &     &                              \\
79$^{g}$ & $9/2^{+}$ &-955.0(6)  &-400.6(4)   &+116.2(50) &-1.1866(10)&+0.487(53)   \\
79$^{m}$& $1/2^{+}$  &-7362.1(61) &-3093.1(36) &      &-1.0180(12)      &           \\
 \hline\hline
\end{tabular}
\footnotetext[1]{Hfs constant $A(^{3}S_{1})$ for the reference $^{67}$Zn measured by laser spectroscopy \cite{Au67Zn}. }
\footnotetext[2]{Hfs constants $A(^{3}P_{2})$, $B(^{3}P_{2})$ for the reference $^{67}$Zn measured by the atomic beam magnetic-resonance technique \cite{Ad67Zn}.}
\footnotetext[3]{Magnetic moment of $^{67}$Zn obtained by the optical pumping technique relative to $^{1}$H \cite{67Zn-moment}.}
\footnotetext[4]{Quadrupole moment of $^{67}$Zn obtained from the optical double resonance measurement and an empirical atomic EFG calculation \cite{Q67Zn}.}
\end{table*}

As pointed out in \mbox{Ref. \cite{Heyde2011}}, the magnetic moment is an important probe of the single-particle nature of nuclear states, while the isomer shift provides information on the relative mean square charge distribution of isomeric states. The combination of both observables can be used to identify and characterize shape coexisting structures.

This Letter reports on the first measurement of nuclear spins and moments of the ground- and long-lived isomeric states in  $^{79}$Zn and more significantly, the substantial increase in charge radius extracted from the isomer shift.

Neutron-rich zinc nuclei were produced in a thick UCx target using 1.4 GeV protons. The $^{79}$Zn nuclei were then resonantly ionized by the Resonant Ionization Laser-Ion Source (RILIS) \cite{RILIS}, was accelerated up to 30 keV and then separated by the high-resolution HRS separator. By using a gas-filled radio-frequency quadrupole ISCOOL \cite{ISCOOL}, the $^{79}$Zn ions were cooled and bunched for typically 200 ms, and then delivered into the collinear laser spectroscopy (COLLAPS) setup \cite{Collaps} as a bunch of 5 $\mu$s temporal length. In order to access a relatively strong transition for efficient spectroscopy, the metastable  $4s4p$ $^{3}P_{2}$ state in the Zn atom was populated via a charge exchange process using sodium vapour, with a neutralisation rate of about 50\%. Laser spectroscopy was performed using a frequency doubled cw Ti:Sa laser, which was locked to 480.7254 nm to match the Doppler shifted \mbox{$4s4p$ $^{3}P_{2}$} $\rightarrow$ \mbox{$4s5s$ $^{3}S_{1}$}  transition. By scanning the voltage applied to the charge exchange cell, the velocity of atoms in the fluorescence detection region was varied, in order to search for a resonant excitation of the hyperfine transitions. The emitted fluorescence photons from the laser-excited atoms were collected by four photomultiplier tubes, and recorded as a function of the scanning voltage to obtain the hyperfine structure (hfs) spectra of the Zn isotopes.

\begin{figure}[!t]
\centering
\vspace{-1mm}
\setbox1=\hbox{\includegraphics[height=5.89cm]{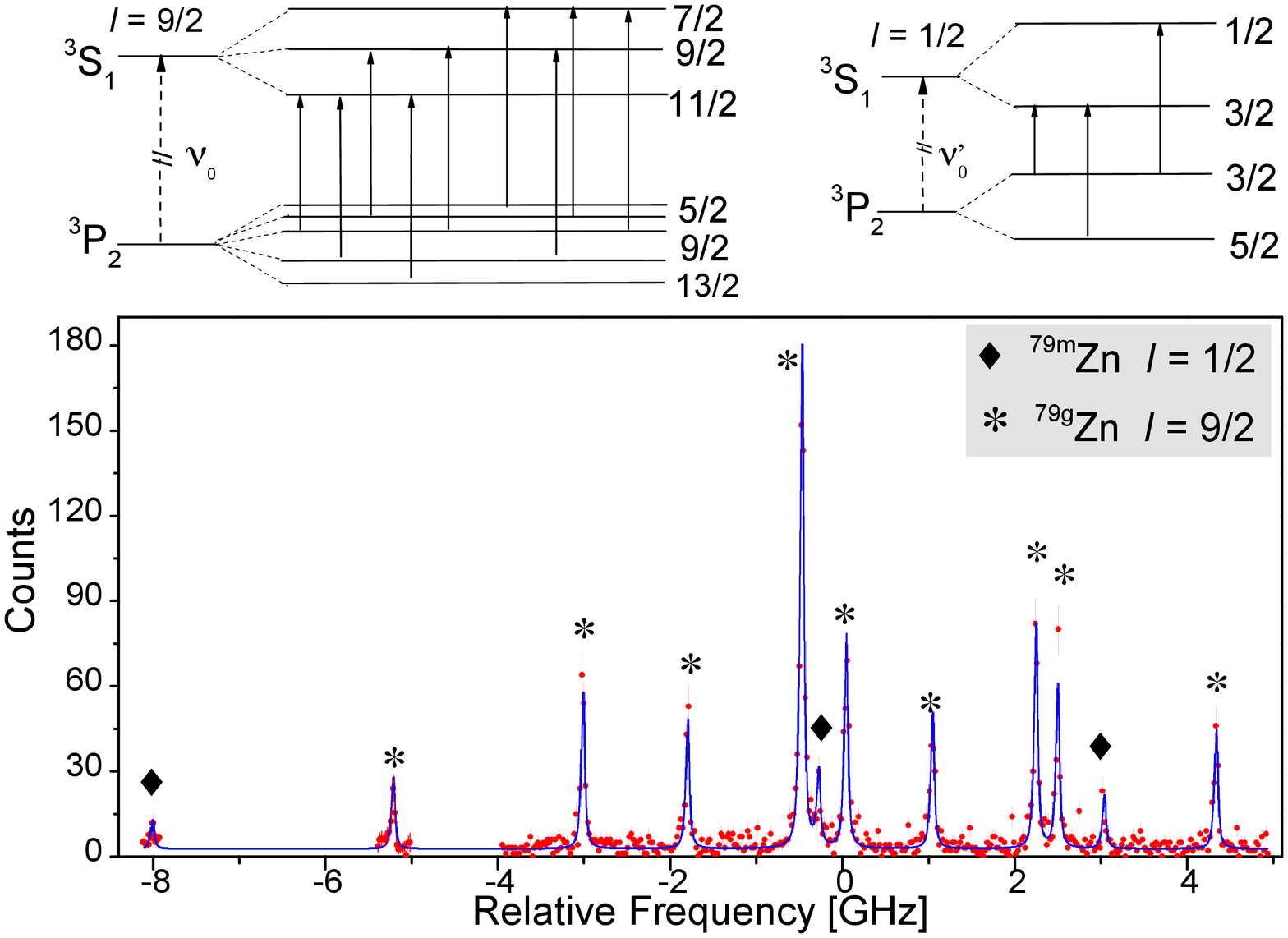}}
\includegraphics[width=.5\textwidth]{Fig1a.eps}\llap{\makebox[\wd1][l]{\raisebox{2.08cm}{\includegraphics[height=2.25cm]{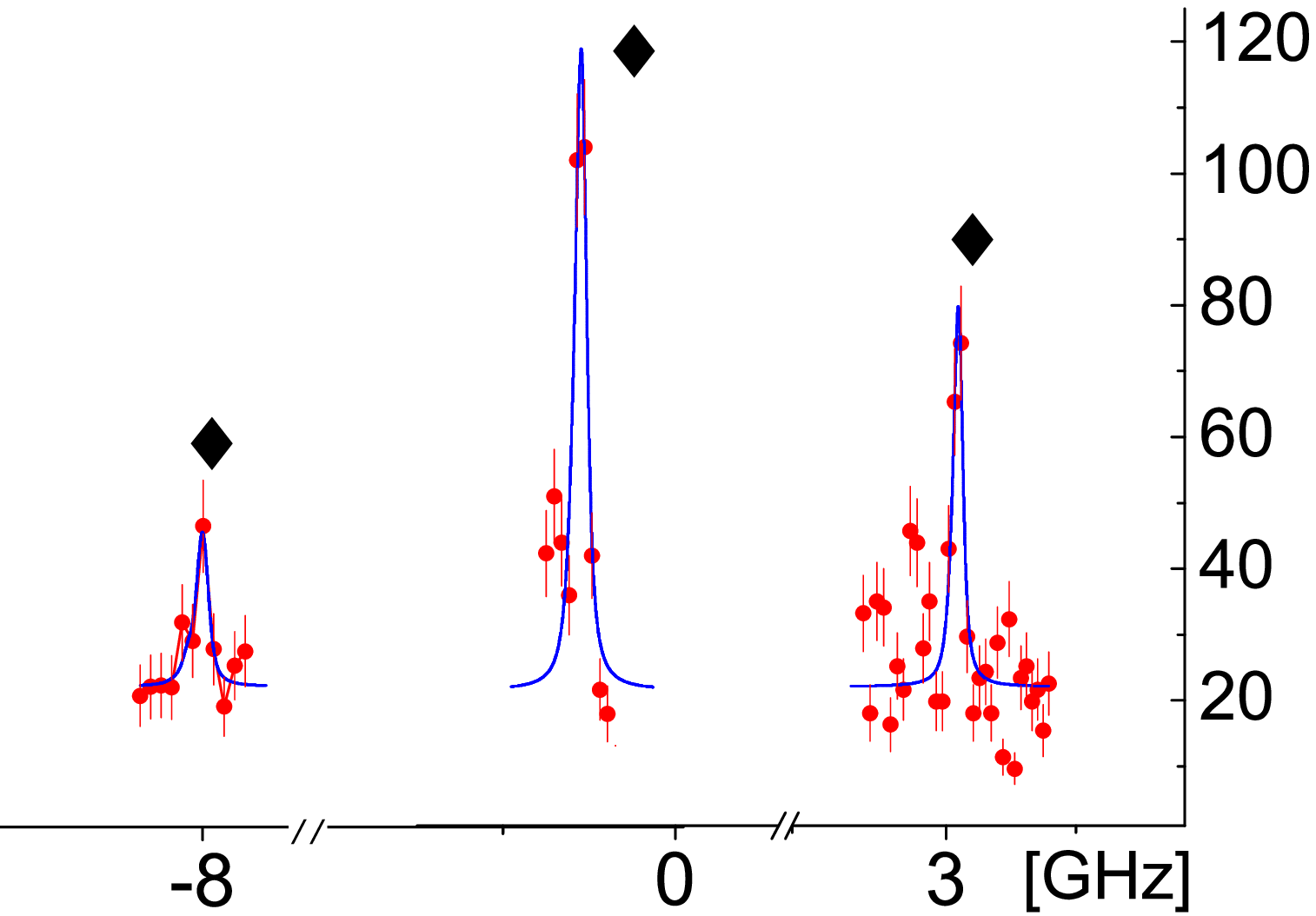}}}}
\vspace{-6mm}
\caption{\footnotesize{(Color online) Typical hyperfine spectra of the ground and
isomeric states in $^{79}$Zn, denoted by asterisks and diamonds, respectively. The three isomeric transitions were confirmed in separate scans around each of the three peaks, as highlighted in the insert. The blue solid lines show the fit with a Lorentzian line profile.}}\label{spectra}
\label{chargeradii}
\vspace{-9mm}
\end{figure}

Typical hfs spectra for $^{79,79m}$Zn are shown in \mbox{Fig. \ref{spectra}}. For a state of nuclear spin $I$ = $1/2$, there are only three hyperfine transitions, as illustrated in the top of \mbox{Fig. \ref{spectra}}. The hyperfine spectra of state with spin $I \geq 5/2$ will contain nine resonance peaks. All the resonance peaks observed for the ground state of $^{79}$Zn, which has a suggested spin of 9/2, are denoted by asterisks. The three resonance peaks denoted by diamonds correspond to the hfs spectrum of the isomer $^{79m}$Zn. These resonances have been further confirmed by separated scans around each resonance, as shown in the insert of \mbox{Fig. \ref{spectra}}. Therefore, spin $1/2$ can be firmly assigned to this isomeric state, which was observed before in a transfer reaction experiment \cite{Orlandi2015}. No measurement or estimation of the half-life for this isomer in $^{79}$Zn has been reported. The observation of the hfs spectrum of the isomeric state in this work establishes a long-lived nature of the isomer for the first time. Indeed, we have measured the ratio of the strongest resonance peaks in the two hfs spectra for different accumulation times in the RFQ (50 ms, 100 ms and 200 ms) after the proton trigger. No significant change in the intensity ratio has been observed, which suggests that the isomeric half-life must be more than 200 ms, as is the lifetime of the ground state \mbox{(746 ms \cite{HosmerPRC2010})}.

The positions of resonance peaks in the spectra of \mbox{Fig. \ref{spectra}} are related to the hyperfine structure constants, the nuclear spin, and the centroid of the transition. Using a $\chi^{2}$  minimization routine (MINUIT), the resonance peaks in the hfs spectra of both ground and isomeric states were fitted with Lorentzian line profiles (blue solid line in \mbox{Fig. \ref{spectra}}) following the procedure described in \mbox{Ref. \cite{laser2015}}. In the fitting procedure, different spins were assumed for the ground state of $^{79}$Zn, but only spin $9/2$ can reproduce all resonance peaks with a reasonable reduced $\chi^{2}$ ($\sim$1.2), thus firmly establishing $I$ = 9/2 for this state. The hfs constants for both atomic states $A(^{3}P_{2} )$, $B(^{3}P_{2})$ and $A(^{3}S_{1})$, the centroid of the transitions and the nuclear spins were extracted from the fitting, as listed in \mbox{Table I}. The $B$ constant for the state $^{3}S_{1}$, also a free parameter in the fit, is smaller than the typical experimental uncertainty (few MHz). Our results for the reference isotope $^{67}$Zn are in good agreement with \mbox{Refs. \cite{Au67Zn,Ad67Zn}}. The magnetic moments $\mu$ of $^{79,79m}$Zn were extracted using the relation $\mu= \mu_{\rm{ref}}IA/(I_{\rm{ref}}A_{\rm{ref}}$) and $^{67}$Zn as the reference isotope, which are presented in \mbox{Table I}. From the measured $B(^{3}P_{2})$ values, quadrupole moments are extracted using the equation of $Q_s= Q_{\rm{ref}}B/B_{\rm{ref}}$ and the $^{67}$Zn as the reference isotope (Table I).

\begin{figure}[!b]
\begin{center}
\vspace{-8mm}
\includegraphics[width=.42\textwidth]{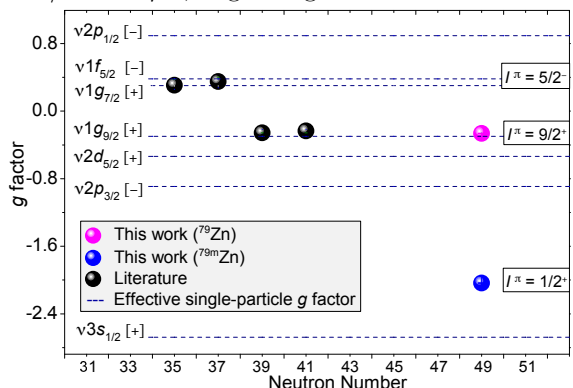}
\vspace{-3mm}
\caption{\footnotesize{(Color online) Measured $g$-factors of the ground state of $^{79}$Zn and its isomer $^{79m}$Zn together with $g$-factors of $^{65,67,69m,71m}$Zn from Refs. \cite{65Zn-moment,67Zn-moment,6971Zn-moment}. Effective single-particle $g$-factors (dashed lines) are calculated for each orbit, of which the parity is marked in the square brackets.}}\label{moment}
\label{gfactor}
\vspace{-7mm}
\end{center}
\end{figure}
Nuclear magnetic moments, and in particular the related $g$-factors \mbox{($g$ = $\mu/I$)}, are a sensitive probe of the configuration of the wave function in near-magic isotopes. In order to study the occupied orbit of the unpaired valence-neutron particle/hole, experimental $g$-factors of Zn isotopes and isomers \cite{65Zn-moment,67Zn-moment,6971Zn-moment} are compared to effective \mbox{single-particle} \mbox{$g$-factors} in \mbox{Fig. \ref{gfactor}}. The effective single-particle $g$-factors (dashed line in \mbox{Fig. \ref{gfactor}}) for each orbit are calculated using $g_{s}$= $-$2.6782 ($g^{\rm{eff}}$  = 0.7$g^{\rm{free}}$) and $g_{l} = 0$, which are the typical values used for the $pfg$ shell \cite{Homma2009}.  From $^{65}$Zn $\rightarrow$ $^{79}$Zn, the unpaired valence neutrons are expected to gradually fill the $\nu 1f_{5/2}$, $\nu 2p_{1/2}$, $\nu 1g_{g/2}$ orbits. The isotopes $^{65,67}$Zn have $g$-factors close to the effective \mbox{single-particle} \mbox{$g$-factor} of the $\nu 1f_{5/2}$ orbit, suggesting a $\nu 1f_{5/2}^n$ configuration for both 5/2$^{-}$ ground states. The isomeric state $g$-factors of $^{69m,71m}$Zn with spin $9/2^{+}$ are close to the effective $g$-factor value for an unpaired neutron in the $\nu 1g_{9/2}$ orbit.

With a single hole in the $\nu 1g_{9/2}$ orbit  with respect to the closed $N$ = 50 shell gap, the ground-state $g$-factor of $^{79}$Zn is close to its effective single-particle $g$-factor. This is also closely reproduced by large scale shell-model calculations using jj44b/JUN45 ($^{56}$Ni core and $pf_{5/2}g_{9/2}$ shell) effective interactions \cite{ChealGa,Homma2009}, yielding $\mu_{\text{jj44b/JUN45}}$ = $-$1.173/$-$1.185 $\mu_{\rm{N}}$, in good agreement with the observed value $-$1.1866(10) $\mu_{\rm{N}}$. The large negative $g$-factor for the \mbox{$I$ = 1/2} isomer $^{79m}$Zn excludes the  $\nu 2p_{1/2}$ as a possible configuration (upper dashed line in \mbox{Fig. \ref{gfactor}}). Looking at the single particle g-factors for all relevant neutron orbits below and above the $N$ = 50 shell gap (\mbox{Fig. \ref{gfactor}}), only the positive parity $\nu 3s_{1/2}$ orbit has a strong enough negative value to account for the experimental number. This supports the earlier proposed positive parity of the $1/2$ isomer in $^{79}$Zn \cite{Orlandi2015}. The deviation of the experimental g-factor from the single particle $\nu 3s_{1/2}$ value suggests some mixing with a configuration where the neutron is excited into the other positive-parity $\nu 2d_{5/2}$ orbit (where it couples to a $2^{+}$ proton or neutron configuration to form a 1/2$^{+}$ state). Thus the experimental magnetic moment and $g$-factor are consistent with a 2h-1p intruder configuration with a dominant $\nu(1g^{-2}_{9/2}3s^{1}_{1/2})[1/2^{+}]$ component and some mixing with e.g. a $\nu(1g^{-2}_{9/2}2d^{1}_{5/2})[1/2^{+}]$ configuration. Note that also 4h-3p configurations with unpaired particles in these same orbits would lead to a similar magnetic moment and such mp-mh configurations can therefore not be excluded. To quantify the mixing and the amount of mp-mh configurations in the wave function, further development of larger scale state-of-the-art shell-model calculations, including not only the $\nu 2d_{5/2}$ \cite{Sieja2012,Tsunoda2014} but also the $\nu 3s_{1/2}$ orbit, is needed.

A remarkable result from this work is the large isomer shift observed for $^{79m}$Zn, which proves a large increase of the rms charge radius with respect to its ground state. From the fitted centroid of the spectra in \mbox{Fig. \ref{spectra}}, the isomer shift $\delta\nu^{79,79m}$ = 61.3(31) MHz, between the ground and isomeric state, is obtained. To extract the change in the rms charge radii of $^{79,79m}$Zn, the field-shift and mass-shift factors, \mbox{$F$ = +301(51) MHz/fm$^{2}$} and \mbox{$K_{MS}$ = +59(18) GHz u}, are determined from a King plot process using the known rms charge radii of stable isotopes ($^{64,66-68,70}$Zn) from an independent measurement \cite{Au67Zn, radiibook}. The extracted difference in the rms radii is $\langle r_{c}^{2}\rangle(^{79m}$Zn) $-$ $\langle r_{c}^{2}\rangle(^{79g}$Zn) = 0.204(6)[36] fm$^{2}$. The error in the square brackets corresponds to the systematic contribution associated with the uncertainty on the atomic factors. A similarly large increase has been observed previously in the rms charge radii difference between the Thallium ground states and their intruder isomeric states \cite{Tl}. In this region the phenomenon of shape coexistence has been well established. Note that a ten times smaller rms increase of only $\sim$0.020(5) fm$^{2}$ has been observed between the ground states of the $N$ = 49 isotones $^{85}$Kr$_{36}$ and $^{87}$Sr$_{38}$ and their normal 1/2$^{-}$ isomeric states \cite{Sr1990,Keim1995}.

\begin{figure}[!t]
\centering
\vspace{-1mm}
\includegraphics[width=.4\textwidth]{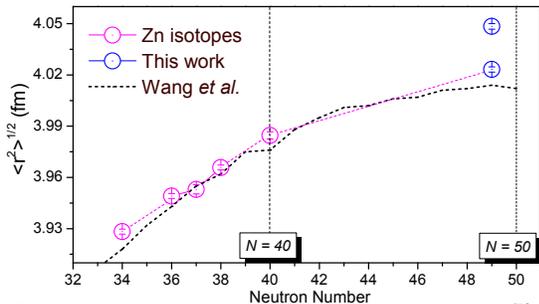}
\vspace{-4.0mm}
\caption{\footnotesize{(Color online) Root mean square charge radii of $^{79,79m}$Zn obtained from this work, together with earlier
data on other Zn \cite{Znradii} isotopes and calculated rms charge radii using a modern phenomenological model by Wang and Li  \cite{Radiical}.}}
\label{chargeradii}
\vspace{-7mm}
\end{figure}
The rms charge radii of Zn isotopes (from \cite{Znradii} and this work) are plotted in \mbox{Fig. \ref{chargeradii}}, along with calculated rms charge radii by Wang and Li \cite{Radiical}. The rms charge radius of the ground state of $^{79}$Zn follows the general trend predicted by this phenomenological model. However, a notably larger rms charge radius is observed for the intruder isomeric state, pointing to a larger deformation. The $^{79}$Zn ground state deformation can be deduced from its measured quadrupole moment (\mbox{Table I)}. Assuming a strong coupling with $\Omega$ = $I$ in \mbox{Eq. 20} of \mbox{Ref. \cite{Laser2010}}, we find $\beta_2$ = 0.15(2). Although it could be accidental, we notice that the same deformation has been deduced from the measured $B(E2 \uparrow)$ transition probabilities in the even-even neighboring isotopes $^{78,80}$Zn: \mbox{$\langle\beta_2^{2}\rangle^{1/2}$ = 0.15(2)} and \mbox{$\langle\beta_2^{2}\rangle^{1/2}$ = 0.14(2)} \cite{Van2009}. Using the deduced ground state deformation, we can now estimate the deformation of the isomeric state, by assuming that its increased charge radius is due to an increase in deformation, using equations (17) and (18) from Ref. \cite{laser2015}. Neglecting a possible increased mean square spherical radius, we find that the larger rms charge radius of the 1/2$^{+}$ isomer corresponds to an increased deformation of $\langle\beta_2^{2}\rangle^{1/2} \sim 0.22$.

This larger deformation can be interpreted as due to the intruder (mp-mh) nature of the isomeric state. The gain in correlation energy in these mp-mh excitations across the $N$ = 50 shell gap strongly reduces the energy of the intruder isomeric states \cite{Heyde2011,PC}. Due to excitations of neutrons across the $N$ = 50 gap, the additional holes in the $\nu g_{9/2}$ orbital drive the intruder state into deformation. Therefore this result provides a first signature of shape coexistence at low energy in $^{79}$Zn.

\begin{figure}[!b]
\begin{center}
\vspace{-8mm}
\subfloat[]{
\label{energy1}\includegraphics[width=.43\textwidth]{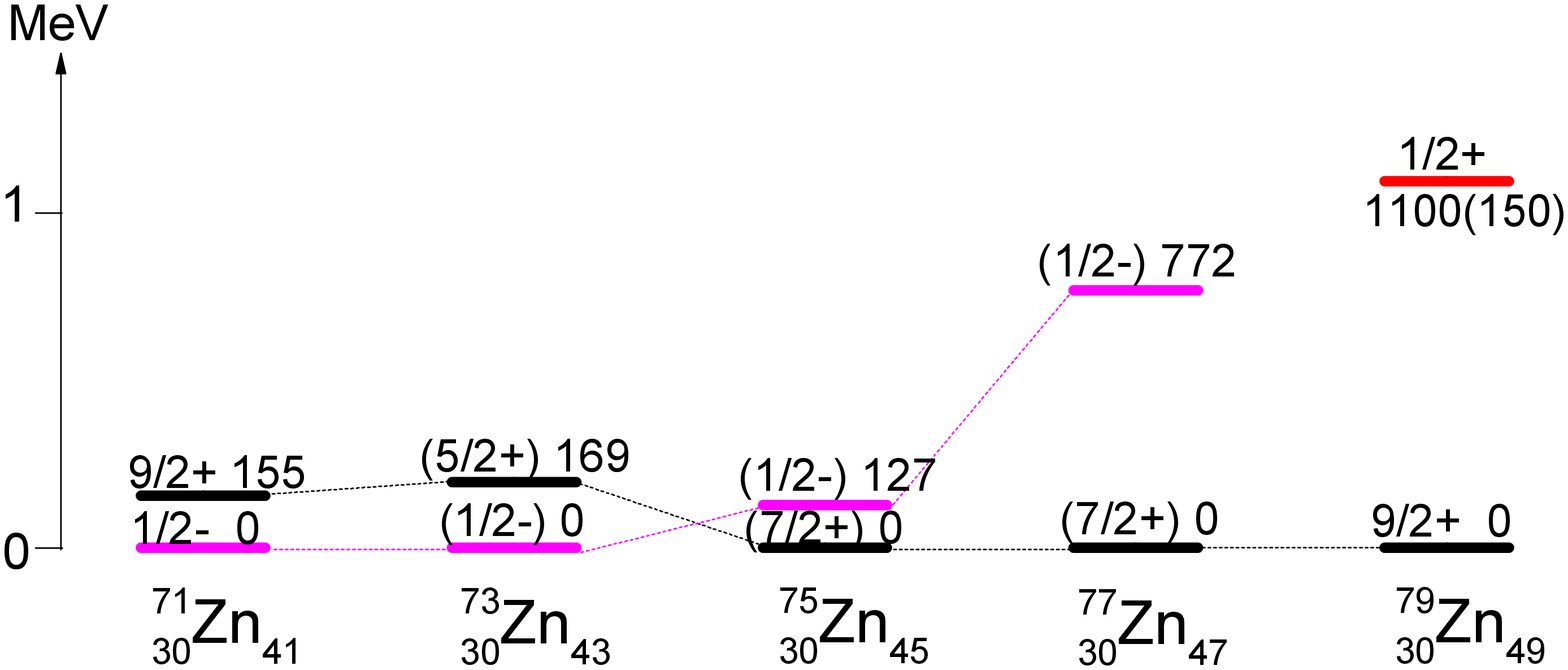}}
\vspace{-5mm}
\subfloat[]{
\label{energy}\includegraphics[width=.43\textwidth]{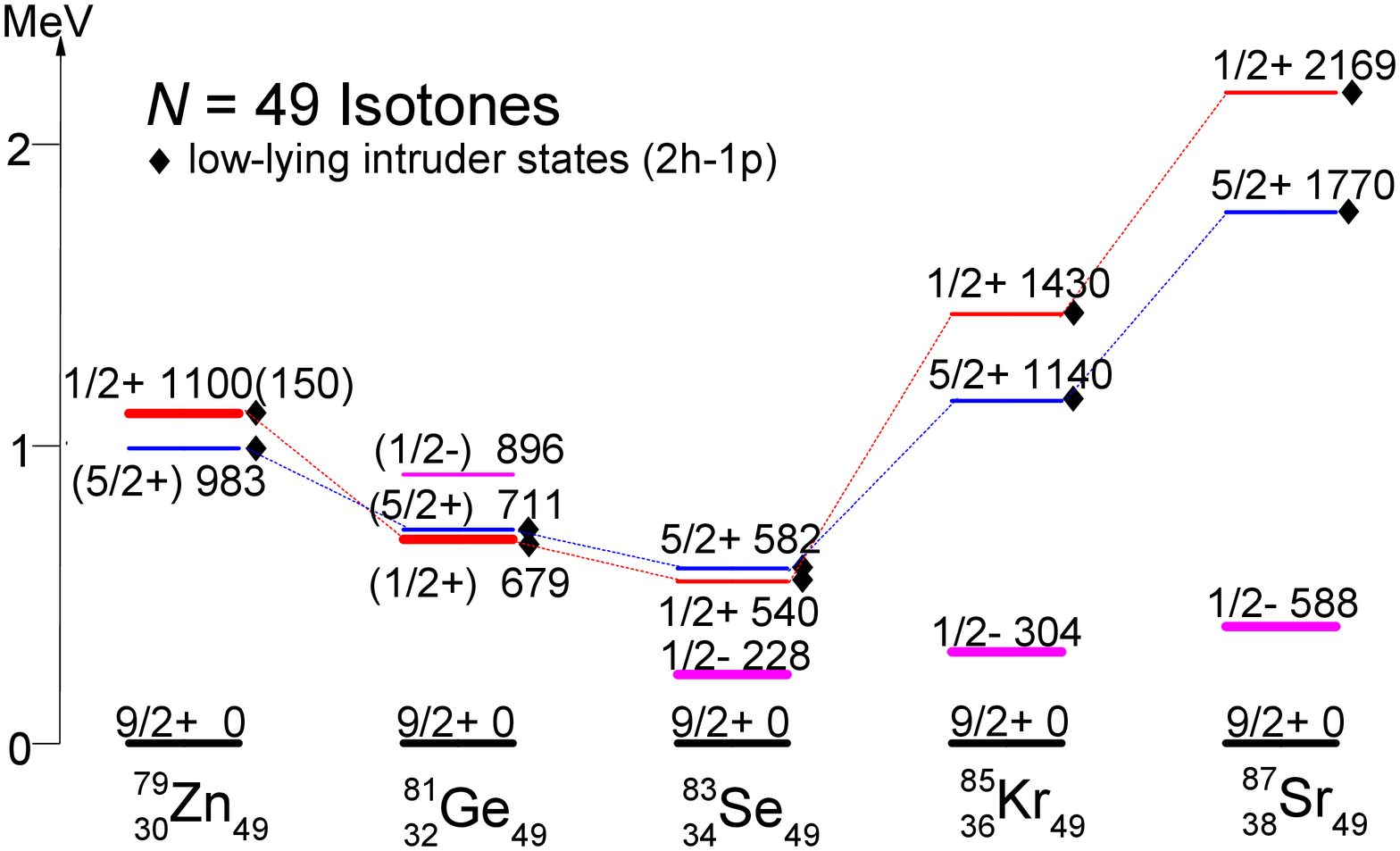}}
\vspace{-4mm}
\caption{\footnotesize{(Color online) (a) Ground and isomeric states in Zn isotopes suggested in the Refs. \cite{Li1967, Huhta1998,Ilyushkin2011,Ilyushkin2009, Orlandi2015}. (b) Odd-mass $N = 49$ level systematics, partially adopted from \mbox{Refs. \cite{Heyde1983,Hoff1981}}. The levels with a thick solid line are the long lived isomeric states, and the diamonds mark the 1/2$^{+}$ and 5/2$^{+}$ intruder states. The firm spin assignment for the $^{79}$Zn levels are from this work.}}
\vspace{-4mm}
\end{center}
\end{figure}

Considering that $^{79}$Zn is only 2 protons above $^{78}$Ni, the energy systematics of the isomeric levels in the Zn isotopic chain (\mbox{Fig. \ref{energy1}} from Refs. \cite{Li1967,Huhta1998,Ilyushkin2011,Ilyushkin2009, Orlandi2015}) as well as along the $N$ = 49 isotones (\mbox{Fig. \ref{energy}} from Refs. \cite{Heyde1983,Hoff1981,Orlandi2015}) is instructive to study the possible \lq magic' nature of $^{78}$Ni. The spin/parity for both ground- and isomeric levels in the Zn isotopes are only tentatively assigned from $^{73}$Zn onwards.  Note that the ground states of the isotopes beyond $N$ = 40 are suggested not to have $I^{\pi}$ = 9/2$^{+}$, which would be expected in a naive shell model picture with a normal filling of the orbits. Only for $^{79}$Zn, the ground state spin/parity is 9/2$^{+}$, as confirmed here. Thus only the one-neutron-hole configuration with respect to $N$ = 50 behaves like a single-particle. The 1/2$^{-}$ state, proposed as the ground state in $^{71,73}$Zn, appears in the heavier isotopes up to $N$ = 47 as a long-lived isomeric state. In $^{79}$Zn, the 1/2$^{-}$ level appears above the positive-parity levels, and the lower-lying intruder level with spin 1/2$^{+}$ becomes isomeric \cite{Orlandi2015}, with a lifetime of several hundreds of milliseconds as discussed above.

Note that there is a rather large uncertainty on the energy of the 1/2$^{+}$ isomer, indicating that the isomer might possibly be located below the 5/2$^{+}$ state and thus below 1 MeV, as is the case in $^{81}$Ge and $^{83}$Se (\mbox{Fig. \ref{energy}}). In such a case, from the Weisskopf estimate, a few hundred ms half-life would indeed be expected for an $E4$ transition ($1/2^{+} \rightarrow 9/2^{+}$), consistent with our observed lifetime. Conversely, if the isomer is located above the 5/2$^{+}$ state, based on the Weisskopf estimate, it should be less than 15 keV above this level (so still below 1 MeV) to obtain an $E2$ transition ($1/2^{+} \rightarrow 5/2^{+}$) that leads to a few hundreds ms isomer. 

This unexpected slow increase of the intruder state energy from the minimum at $Z$ = 34 to less than 1 MeV at $Z$ = 30, contrary to the steep increase towards more than 2 MeV at $Z$ =38, may challenge the magicity of the \lq\lq double magic" isotope, $^{78}$Ni.

In summary, the hfs of the ground and isomeric states of $^{79}$Zn were measured for the first time. The data confirm the long-lived isomer with a few hundreds milliseconds half-life, and establish the spin-parity $9/2^{+}$ and $1/2^{+}$ for the ground and isomeric states. The intruder nature of this isomer has been revealed from its $g$-factor, which is closer to the neutron single-particle value of $\nu 3s_{1/2}$ orbital. While the $g$-factor of the $^{79}$Zn ground state is well accounted for by shell-model calculations including only the $pf_{5/2}g_{9/2}$ neutron orbits, a larger model space is required to account for the isomeric state g-factor. Furthermore, the large rms charge radius of the isomer relative to that of the ground state is linked to a larger deformation of the isomer, providing evidence for shape coexistence in $^{79}$Zn. The slow increase of the intruder state energy towards $Z$ = 28, together with the unexpected appearance of shape coexistence near a supposedly \lq doubly-magic' $^{78}$Ni, challenges the magicity of $^{78}$Ni. Further theoretical and experimental investigations are encouraged in this region. For example, Coulomb excitation of the isomeric state could give complementary information on its deformation. A measurement of the half-life and the exact energy of the 1/2$^{+}$ state in $^{79}$Zn, the measurement of the 1/2$^{+}$ isomeric state properties in $^{81}$Ge, and a study of the low-lying level properties in $^{77}$Ni will help to further enhance our understanding of 2h-1p intruder states induced by neutron excitations across \mbox{$N$ = 50}.
\begin{acknowledgments}
The authors would like to thank \mbox{Prof. J. Wood} and \mbox{Prof. K. Heyde} for valuable discussions. The support and assistance from the ISOLDE technical group are gratefully acknowledged. This work was supported by the IAP-project P7/12,the FWO-Vlaanderen, GOA grant 15/010 from KU Leuven, the NSF grant PHY-1068217, the BMBF Contracts Nos. 05P15RDCIA, the Max-Planck Society, the Science and Technology Facilities Council, and the EU FP7 via ENSAR No. 262010.
\end{acknowledgments}
\bibliography{references}

\end{document}